\documentclass[aps,pra,twocolumn,superscriptaddress]{revtex4-1}
\usepackage[utf8]{inputenc}
\usepackage{graphicx}   
\usepackage{amssymb}
\usepackage{amsmath}
\usepackage{xcolor}
\usepackage{bm}

\begin{document}

\title{Spin and mass currents near a moving magnetic obstacle in a two-component Bose-Einstein condensate}

\author{Jong Heum Jung}
\affiliation{Department of Physics and Astronomy, Seoul National University, Seoul 08826, Korea}

\author{Hyung Jin Kim}
\affiliation{Department of Physics and Astronomy, Seoul National University, Seoul 08826, Korea}

\author{Y. Shin}
\email{yishin@snu.ac.kr}
\affiliation{Department of Physics and Astronomy, Seoul National University, Seoul 08826, Korea}
\affiliation{Institute of Applied Physics, Seoul National University, Seoul 08826, Korea}

\date{\today}

\begin{abstract}
We study the spatial distributions of the spin and mass currents generated by a moving Gaussian magnetic obstacle in a symmetric, two-component Bose-Einstein condensate in two dimensions. We analytically describe the current distributions for a slow obstacle and show that the spin and the mass currents exhibit characteristic spatial structures resembling those of electromagnetic fields around dipole moments. When the obstacle's velocity increases, we numerically observe that the flow pattern maintains its overall structure while the spin polarization induced by the obstacle is enhanced with an increased spin current. We investigate the critical velocity of the magnetic obstacle based on the local criterion of Landau energetic instability and find that it decreases almost linearly as the magnitude of the obstacle's potential increases, which can be directly tested in current experiments.
\end{abstract}



\maketitle

\section{INTRODUCTION}

A superfluid can flow without fricition, but its superfluidity breaks down above a certain critical velocity. The critical velocity is mainly determined by the intrinsic excitation properties of the superfluid~\cite{R1}, and its manifestation is significantly affected by the geometry and boundary condition of the flowing channel. Understanding the critical dynamics which involves energy dissipation processes is important in the study of a superfluid system. For ultracold atomic gas experiments, a simple method was developed to investigate the critical velocity of a superfluid. In that method, a sample is stirred with an optical obstacle formed by focusing a laser beam, and the onset of dissipation due to the increase in the obstacle velocity is detected via the increase in the sample temperature~\cite{R2,Dalibard12,R5} or the generation of topological defects such as quantized vortices~\cite{R7,Neely10,Kwon15a,Park18}.  Finite critical velocities were presented as evidence for superfluidity~\cite{R2,Dalibard12}, and the measured values provided quantitative tests for our microscopic understanding of superfluid systems~\cite{R5,Kwon15a,Park18,R6}.

Recently, a symmetric binary superfluid gas system was experimentally realized using a Bose-Einstein condensate (BEC) of $^{23}$Na in two hyperfine spin states, i.e., $|F=1,m_F=1\rangle$ and $|F=1,m_F=-1\rangle$~\cite{R20}. This system, with a $\mathrm{Z}_2$ symmetry, constitutes a minimal setting for studying superfluidity with multiple order parameters. Spin superfluidity was demonstrated with the absence of damping in spin dipole oscillations in trapped samples~\cite{R16,R19}, and novel topological objects such as half-quantum vortices~\cite{R20,R21} and magnetic solitons~\cite{R24,R25} were observed. These developments lead us to anticipate a moving obstacle experiment with the binary superfluid system, discussed in previous numerical studies~\cite{R34,R35,Kamchatnov13,R36}. In particular, the optical obstacle can be engineered to be magnetic, i.e., exhibiting different potentials for the two spin components so that the system's properties in both the spin and the mass sectors may be addressed in a controlled manner. Considering different topological objects and peculiar dynamic effects such as countersuperflow instability~\cite{R29,R27}, such an experiment may open a way to investigate a new class of critical superfluid dynamics~\cite{Kamchatnov13}. In a recent experiment, a localized spin-dependent optical potential was indeed used to measure the speed of spin sound in a binary $^{23}$Na BEC~\cite{R26}. Therefore, the stirring experiment with a magnetic obstacle is within immediate reach.

Herein, we theoretically consider a primary case in which a penetrable, Gaussian magnetic obstacle moves in a symmetric binary BEC in two dimensions (2D). Based on the hydrodynamic equations of the two-component BEC system, we analytically and numerically investigate the spatial distributions of the induced superflows around the moving obstacle and demonstrate that the spin and the mass supercurrents are formed in characteristic spatial structures resembling those of electric and magnetic fields around a charge dipole and a current loop, respectively. Furthermore, we investigate the local Landau instability of the induced superflows and determine the critical velocity, $u_c$, of the magnetic obstacle as a function of its potential magnitude, $V_0$. We find that $u_c$ decreases almost linearly from the speed of spin sound with increasing $V_0$, which can be directly tested in current experiments. This study provides a basis for the study of the critical dynamics of binary superfluid systems with a moving magnetic obstacle.

The remainder of this paper is organized as follows: In Section~II, we present a hydrodynamic description of the spin and the mass currents near a moving magnetic obstacle in a two-component BEC. In Section~III, we first analyze the characteristic superflow pattern in a slow obstacle limit and then numerically investigate the evolution of the current distributions with increasing obstacle velocity. The critical velocity of the magnetic obstacle is determined by examining the local speed of sound at the obstacle and applying the Landau criterion. Finally, in Section~IV, we provide a summary and some outlooks on future experimental studies.

\section{Hydrodynamic model}

Figure~1 shows the physical situation of our interest, where a localized Gaussian potential traverses a homogeneous two-component BEC in 2D with a constant velocity, $\bm{u}=u\hat{\bm{x}}$. The BEC is a balanced mixture of two miscible components denoted by spin-$\uparrow$ and $\downarrow$, separately. They are identical to each other in terms of particle mass and intracomponent interactions, and the BEC represents a symmetric binary superfluid system. The Gaussian potential is spin dependent, which is attractive to the spin-$\uparrow$ component and repulsive to the spin-$\downarrow$ component, i.e., $V_{\uparrow(\downarrow)}(\bm{r})=-s_{\uparrow(\downarrow)} V(r)$ with $s_\uparrow=-s_\downarrow=1$ and $V(r)= V_0 \exp(-\frac{2 r^2}{\sigma^2})$. As such, the moving magnetic potential will generate different flow patterns for the two spin components. The main focus of this study is to investigate the spin and the mass currents near the moving obstacle; these are defined as $\bm{J}=n_\uparrow \bm{u}_\uparrow - n_\downarrow \bm{u}_\downarrow$ and $\bm{M}=n_\uparrow \bm{u}_\uparrow + n_\downarrow \bm{u}_\downarrow$, respectively, with $\{ n_{\uparrow(\downarrow)}(\bm{r},t), \bm{u}_{\uparrow(\downarrow)}(\bm{r},t) \}$ being the density distribution and the velocity field of the spin-$\uparrow(\downarrow)$ component, respectively.

\begin{figure}[t!]
    \includegraphics[width=8.4cm]{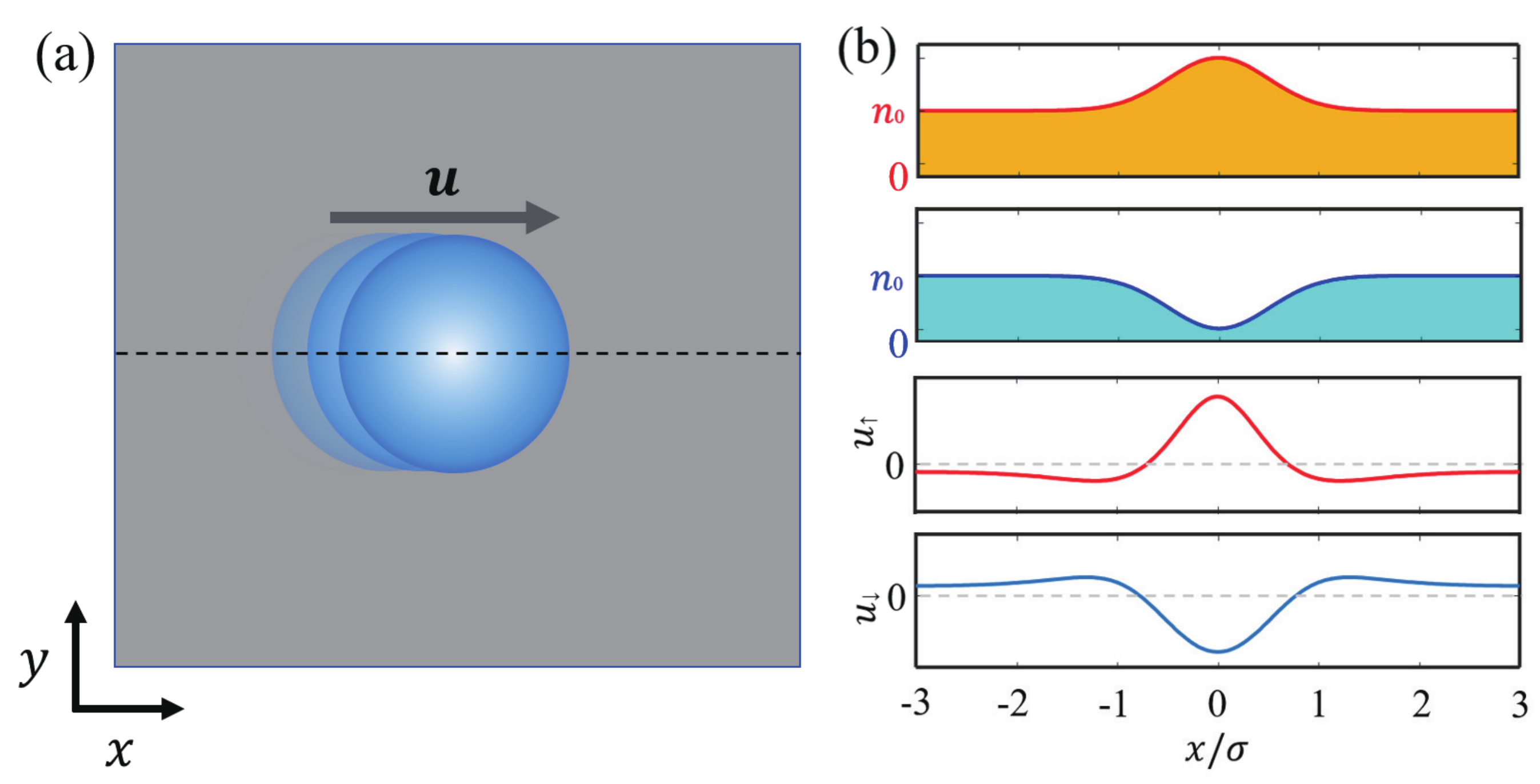}
\caption{Generation of supercurrents by a moving magnetic obstacle in a two-component Bose-Einstein condensate (BEC). (a) A penetrable Gaussian obstacle (blue) moves with a constant velocity $\bm{u}$ in a homogeneous two-dimensional BEC comprising two symmetric, spin-$\uparrow$ and $\downarrow$ components. The obstacle attracts the spin-$\uparrow$ component and repels the spin-$\downarrow$ component, and the two components have different density and velocity field distributions, denoted by $n_{\uparrow(\downarrow)}$ and $\bm{u}_{\uparrow(\downarrow)}$, respectively, near the moving obstacle. (b) Schematic description of the density and the velocity profiles along the horizontal dashed line in (a). $n_0$ is the density of each component for an unperturbed BEC.}
    \label{fig.1}
\end{figure}

In the hydrodynamic approximation, the superflow dynamics of the binary superfluid system can be expressed as follows:
\begin{eqnarray}
\partial_t n_i &+&\nabla \cdot (n_i \bm{u}_i) =0  \\
m\partial_t \bm{u}_i &+& \nabla (\frac{1}{2} m u_i^2 + g n_i + g_{\uparrow\downarrow} n_j) = -\nabla V_i(\bm{r}-\bm{u}t),
\end{eqnarray}
where $i,j=\uparrow,\downarrow$ ($i\neq j$), $m$ is the particle mass, and $g(g_{\uparrow\downarrow})>0$ is the coefficient of the inter(intra)-component interactions. The first equation is the continuity equation for mass conservation, and the second one is the Euler equation associated with energy conservation. The hydrodynamic equations were derived from the Gross--Pitaevskii equations for the wave functions of the BEC, $\psi_i(\bm{r},t)$, under a Madelung transformation of $\psi_i(\bm{r},t)=\sqrt{{n}_i(\bm{r},t)}e^{i \theta_i(\bm{r},t)}$ and $\bm{u}_i (\bm{r},t)=\frac{\hbar}{m} \nabla \theta_i(\bm{r},t)$~\cite{Stringari96,R32}. $\theta_i(\bm{r},t)$ is the macroscopic phase of the spin-$i$ component and $\hbar=\frac{h}{2\pi}$ with the Planck constant $h$. The quantum pressure term is neglected, assuming that the obstacle width, $\sigma$, is much larger than the healing length of the condensate and that the potential magnitude, $V_0$, is small such that $n_i >0$, i.e., no density-depleted region exists for either spin component.

In the co-moving reference frame with the obstacle, the density distributions and the velocity fields are time independent, and the problem becomes more tractable. Under a Galilean transformation $\bm{r}\rightarrow \bm{r}+\bm{u}t$, the hydrodynamic equations reduce to 
\begin{eqnarray}
&&\nabla \cdot (n_i \bm{v}_i) =0, \\
&& \nabla (\frac{1}{2} m v_i^2 - s_i V+ g n_i + g_{\uparrow\downarrow} n_j ) =0,
\end{eqnarray}
with $\bm{v}_i(\bm{r})=\bm{u}_i(\bm{r},0)-\bm{u}$.  
For the boundary conditions of $n_i= n_0$ and $\bm{v}_i = -u\hat{\bm{x}}$ as $r\rightarrow\infty$, Eq.~(4) requires $\frac{1}{2}m{{v}_i}^{2}  - s_i {V} +g {n}_{i}+ {g}_{\downarrow\uparrow} {n}_j =\frac{1}{2}m u^2 +(g+{g}_{\uparrow\downarrow})n_0$, resulting in
\begin{equation}
    n_i = n_0+ \frac{s_i V}{\Delta g}+\frac{m \Big( g (u^2 -v_i^2) -g_{\uparrow\downarrow}(u^2-v_j^2 )\Big)}{2(g^2-g_{\uparrow \downarrow}^2)}, 
\end{equation}
with $\Delta g = g-g_{\uparrow\downarrow}$. Here, $\Delta g>0$ is due to the miscibility condition for the two spin components. From the irrotational property of $\nabla \times \bm{v}_i=0$, a potential function $S_i(\bm{r})$ for $\bm{v}_i(\bm{r})$ can be defined such that $\bm{v}_i=\nabla S_i$, and Eq.~(3) can be rewritten as
\begin{equation}
\nabla^2 S_i +\frac{1}{n_i}\nabla n_i \cdot \nabla S_i =0.
\end{equation}
Once $\{n_i(\bm{r}), S_i(\bm{r})\}$ are determined from Eqs.~(5) and (6), the spin and the mass currents at $t=0$ in the stationary BEC reference frame can be calculated as 
 \begin{eqnarray}
     \bm{J}&=&{n}_{\uparrow}\nabla S_{\uparrow}-{n}_{\downarrow}\nabla S_{\downarrow}+m_z \bm{u} \\ \bm{M}&=&{n}_{\uparrow}\nabla S_{\uparrow}+{n}_{\downarrow}\nabla S_{\downarrow} + n_t \bm{u},
 \end{eqnarray}
respectively, where  $m_z(\bm{r})=n_\uparrow-n_\downarrow$ is the magnetization density, and $n_t(\bm{r})=n_\uparrow+n_\downarrow$ is the total number density of the BEC.

\section{Results}

\subsection{Slow Obstacle}

We first consider a perturbative regime in which the obstacle moves slowly such that the densities of the spin components are well approximated by the solutions of Eq.~(5) for $u=0$, i.e., $n_i(r)=n_0+ s_i V(r)/\Delta g$. In a later discussion, it will be clear that the approximation is valid when the kinetic energy of the induced flow is negligible compared to the spin interaction energy, i.e., $mu^2 \ll m c_s^2 \equiv \Delta g n_0 $. Here, $c_s$ is the speed of spin sound for the unperturbed BEC.

For the density distribution $n_i(r)$, the potential function $S_i(\bm{r})$ can be directly determined using Eq.~(6). Because $n_i$ has only an $r$-dependence, we perform separation of variables, i.e., ${S_i(r,\phi)}={R_i(r)}{\Phi_i(\phi)}$, and Eq.~(6) is transformed to  
\begin{eqnarray}
&&\frac{{d}^{2}{R}_i}{{d}{r}^{2}}+\frac{1}{r}\frac{{d}{R_i}}{{d}{r}}+\frac{1}{1+\delta\tilde{n}_i}\frac{d \delta\tilde{n}_i}{dr}\frac{{d}{R}_i}{{d}{r}}-\frac{{l}^{2}}{{r}^{2}}R_i=0,  \\
&&\frac{{d}^{2}{\Phi_i}}{{d}{\phi}^{2}}+{l}^{2}{\Phi_i}=0, 
\end{eqnarray}
with $\delta\tilde{n}_i(r)=n_i(r)/n_0-1$ and $l$ being an integer. The boundary condition of ${S}_i\rightarrow -u r \cos{\phi}$ as $r\rightarrow \infty$ imposes ${l}=1$, and without loss of generality, we set $\Phi_i(\phi)=-u \cos \phi$. The solution for the radial function, $R_i(r)$, can be obtained perturbatively using the small parameter $\alpha=|\delta\tilde{n}_i(0)|=V_0/(\Delta g n_0)\ll1$, which is the maximum magnitude of the relative density variations in each spin component. When the radial function is expanded in a power series with respect to $\alpha$ as $R_i(r)=\sum_{k\geq 0} \alpha^k R_i^{(k)}(r)$, the $k$-th function $R_i^{(k)}(r)$ is recursively determined from Eq.~(9) as the solution to the following equation: 
\begin{eqnarray}
    \frac{{d}^{2}{R}_{i}^{(k)}}{{d}{r}^{2}}&+&\frac{1}{r}\frac{{d}{R}_{i}^{(k)}}{{d}{r}}-\frac{1}{{r}^{2}}{R}_{i}^{(k)} \nonumber \\
    &&= \frac{4r}{\sigma^2} s_i \sum_{s=1}^{k} {s_i}^{s} e^{-\frac{2sr^2}{\sigma^2}} \frac{dR_{i}^{k-s}}{dr}.
\end{eqnarray}
Up to the second order of $\alpha$, we have
\begin{eqnarray}
R_i^{(0)}(r)&=&r, \nonumber \\
R_i^{(1)}(r)&=& s_i r \frac{-1+{e}^{-2{\rho}^2}}{4\rho^2}, \nonumber \\
R_i^{(2)}(r)&=&r\Big(\frac{1+2{e}^{-2\rho^2}-3{e}^{-4{\rho}^{2}}}{16{\rho^2}}-\frac{1}{4}\int_{2{\rho}^{2}}^{4{\rho}^{2}}{\frac{{e}^{-t}}{t}}dt\Big), \nonumber
\end{eqnarray}
where $\rho=r/\sigma$, thereby yielding the approximate solution of $S_i(\bm{r})$ as  
\begin{equation}
S_i(r,\phi)=- u \big[ R_i^{(0)}+\alpha R_i^{(1)} +\alpha^2 R_i^{(2)}\big] \cos \phi,    
\end{equation}
which satisfies the boundary condition as $r\rightarrow \infty$. 

From Eqs.~(7) and (8) with $n_i(r)=n_0(1+s_i \alpha e^{-2\rho^2})$ and $S_i(r,\phi)$ in Eq.~(12), we obtained the analytic expressions of the spin and the mass currents as
\begin{eqnarray}
    \bm{J}&=&2\alpha{n}_{0}{u}  \Bigg[ \frac{(2\rho^2+1){e}^{-2{\rho}^{2}}-1}{4{\rho}^{2}}(\cos{2\phi}\,\hat{\bm{x}}+\sin{2\phi}\,\hat{\bm{y}})
    + \nonumber \\
    &&+\frac{{e}^{-2{\rho}^{2}}}{2}\,\hat{\bm{x}} \Bigg] \\
    \bm{M}&=&2\alpha^2{n}_{0}{u} \Bigg[ 
    \frac{{({e}^{-2{\rho}^{2}}-1)}^{2}}{16{\rho}^{2}}(\cos{2\phi}\,\hat{\bm{x}}+\sin{2\phi}\,\hat{\bm{y}})+ \nonumber \\
    &&\frac{1}{4}\int_{2{\rho}^{2}}^{4{\rho}^{2}}{\frac{{e}^{-t}}{t}}dt\, \hat{\bm{x}}\Bigg],
\end{eqnarray}
respectively. Of note is that $|\bm{J}|\propto \alpha$ and $|\bm{M}|\propto \alpha^2$, which indicate that the moving magnetic obstacle dominantly generates a spin current, as expected, as well as a mass current via a nonlinear process. The peak spin and mass currents occur at the obstacle center, and are given by $\bm{J}_0=\alpha n_0 \bm{u}$ and $\bm{M}_0=\frac{\ln 2}{2}\alpha^2 n_0 \bm{u}$, respectively.

\begin{figure}[t!]
     \includegraphics[width=8.4cm]{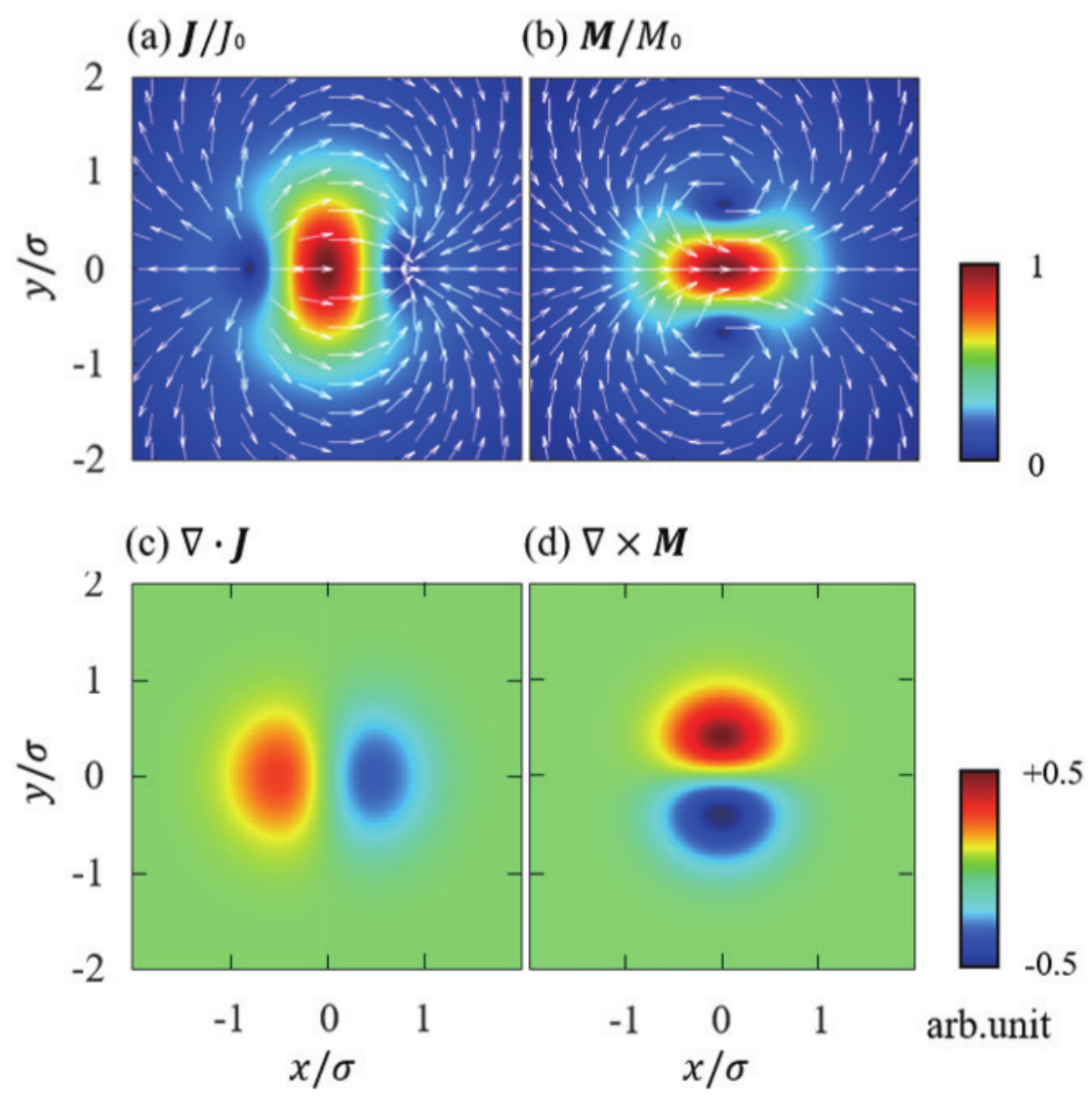}
     \caption{Spin and mass superflows near a moving magnetic obstacle. Spatial distributions of (a)  the spin current $\bm{J}$ and (b) the mass current $\bm{M}$ from the analytical expression of Eqs.~(13) and (14), respectively. The arrow indicates the current's direction and the color denotes the current's magnitude normalized to the peak value $J_0$ ($M_0$) for the spin (mass) current at the obstacle's center. Spatial distributions of (c) $\nabla \cdot \bm{J}$ and (d) $\nabla \times \bm{M}$.}
     \label{fig.2}
\end{figure}

\subsection{Spin and Mass Flow Patterns}
Figures~2(a) and (b) show the spin and the mass current distributions predicted using Eqs.~(13) and (14), respectively. We used $g_{\uparrow\downarrow}/g=0.93$, which is the value for a mixture of $^{23}$Na in the $|F=1,m_F=\pm1\rangle$ states~\cite{R17,Knoop11}. As expected, both spin and mass currents appeared locally near the moving obstacle. In the spin current distribution, we observed two low-current holes, one in the front and the other in the back of the obstacle; furthermore, the spin current flowed out from the back hole and into the front one. Meanwhile, in the mass current distribution, we observed two low-current holes on the lateral sides of the obstacle, and the mass current swirled around each hole in opposite directions. The flow patterns of $\bm{J}$ and $\bm{M}$ around the moving obstacle resembled those of an electric field around a charge dipole and a magnetic field around a current loop, respectively. In Figs.~2(c) and 2(d), we present the distributions of $\nabla\cdot\bm{J}$ and $\nabla\times \bm{M}$, respectively, which clearly show the dipole configurations of the source and the sink for the spin current and the vorticity of the mass current, respectively.

To understand the characteristic flow patterns of the spin and the mass currents, we analyzed the general divergence and rotation properties of $\bm{J}$ and $\bm{M}$. From the continuity equation in Eq.~(3), we obtained $\nabla\cdot (n_i \bm{u}_i )=\bm{u}\cdot \nabla n_i$. Combining the latter with the irrotational property of $\nabla\times\bm{u}_i=0$, we obtain the following relations:
\begin{eqnarray}
    \nabla \cdot \bm{J}&=& \bm{u} \cdot \nabla {m_z}\\
    \nabla \cdot \bm{M} &=& \bm{u} \cdot \nabla{n_t} \nonumber \\
    \nabla \times \bm{J} &=&  \nabla m_z \times \Big( \frac{ \bm{u}_\uparrow +\bm{u}_\downarrow}{2} \Big) + \nabla n_t \times \Big( \frac{ \bm{u}_\uparrow - \bm{u}_\downarrow}{2} \Big) \nonumber \\
    \nabla \times \bm{M} &=&  \nabla m_z \times \Big( \frac{\bm{u}_\uparrow - \bm{u}_\downarrow}{2} \Big)+ \nabla n_t \times \Big( \frac{ \bm{u}_\uparrow + \bm{u}_\downarrow}{2} \Big). \nonumber
\end{eqnarray}
The first and the second equations result from spin and mass conservation, respectively, and the third and the fourth ones reveal intriguing nonlinear coupling between the spin and the mass channels in the binary system.

For a weak and slow magnetic obstacle, taking the same level of approximation as in the previous subsection, we have $\nabla n_t= 0$, $\bm{u}_\uparrow+\bm{u}_\downarrow = \mathcal{O}(\alpha^2)$, and $\bm{u}_\uparrow-\bm{u}_\downarrow=\frac{\bm{J}}{n_0}+\mathcal{O}(\alpha^3)$.  Then, up to the first order in $\alpha$, the relations can be expressed as
\begin{eqnarray}
    \nabla \cdot \bm{J}&=& \bm{u} \cdot \nabla {m_z} \equiv Q_J \nonumber \\
    \nabla \cdot \bm{M} &=& 0 \nonumber \\
    \nabla \times \bm{J} &=& 0 \nonumber \\
    \nabla \times \bm{M} &=&  \nabla m_z \times \frac{\bm{J}}{2n_0} \equiv \bm{I}_M,
\end{eqnarray}
which immediately explains the observed electric- and magnetic-field-like behaviors of $\bm{J}$ and $\bm{M}$, respectively, near the moving magnetic obstacle. The quantities $Q_J$ and $\bm{I}_M$ can be regarded as the `charge' and the `current' source densities for generating the spin and the mass currents, respectively. Their expressions are consistent with the previous results of $|\bm{J}|\propto \alpha$ and $|\bm{M}|\propto \alpha^2$ for $m_z\propto \alpha$. Furthermore, we may consider the `electric' and the `magnetic' dipole moments as 
\begin{eqnarray}
     \bm{p}_{J}&=&\int \bm{r}~Q_J(\bm{r})d^2\bm{r}=\bm{u} \int m_z(r) d^2 \bm{r}  \nonumber \\
     \bm{\mu}_{M}&=&\frac{1}{2}\int \bm{r}\times\bm{I}_M(\bm{r}) d^2\bm{r}=\frac{1}{2 n_t}\int m_z(\bm{r})\bm{J}(\bm{r}) d^2\bm{r}, \nonumber
\end{eqnarray}
respectively, which allow us to predict the currents in the far distant region of $r\gg \sigma$ to be $\bm{J}\sim\frac{2(\bm{p}_{J}\cdot\hat{r})\hat{r}-\bm{p}_{J}}{2\pi r^2}$ and $\bm{M} \sim \frac{2(\bm{\mu}_{M}\cdot\hat{r})\hat{r}-\bm{\mu}_{M}}{2\pi r^2}$.  We emphasize that the relations in Eq.~(16) hold regardless of the potential form of $V(\bm{r})$ once the magnetic obstacle is weak and slow.

The existence of nonzero $\nabla \times \bm{M}$ should be highlighted. The superfluid velocity of the binary superfluid system can be expressed as ${\bm{u}}_{M}=\frac{\bm{M}}{n_t}$; therefore, the circulation of ${\bm{u}_M}$ can be nonzero for $\nabla \times \bm{M}\neq 0$. This is in stark contrast to the case with a single-component BEC, where the circulation of the superfluid velocity should be quantized with $h/m$ as a topological invariant of the system. Noting that the mass circulation of the binary superfluid system can have a continuous value in conjunction with the spin current is important.

\begin{figure}[t!]
\includegraphics[width=8.4cm]{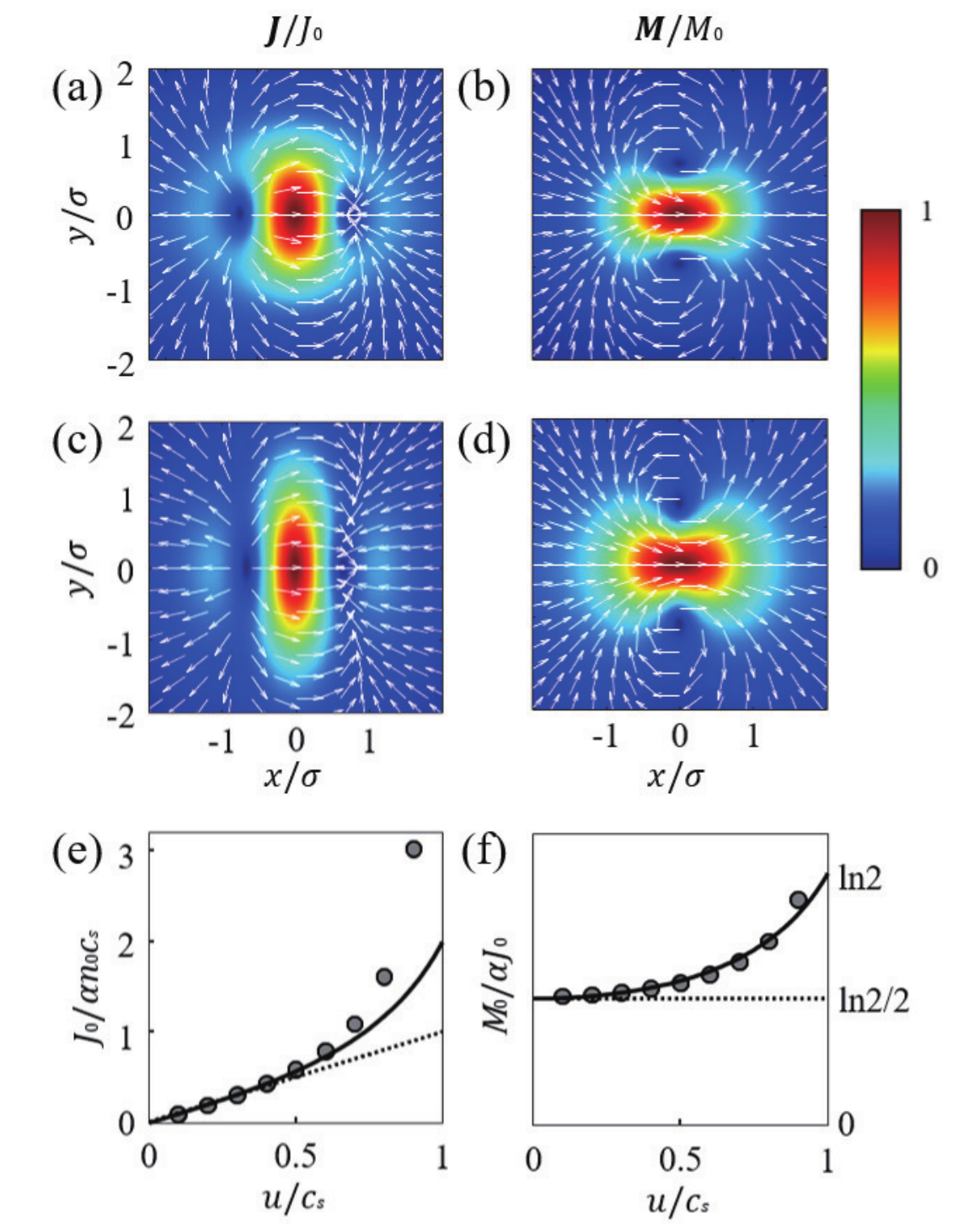}
\caption{Numerical results for the spin and the mass current distributions for (a, b) $u=0.1 c_s$ and (c, d) $u=0.9 c_s$. Here $c_s$ is the speed of spin sound in the unperturbed BEC. In the numerical simulations, $\Delta g/g=0.07$ and $V_0=0.1 m c_s^2$. (e) $J_0$ and (f) $M_0/J_0$ as functions of $u$. $\alpha\equiv V_0/(m c_s^2)=0.1$. The dotted lines indicate the analytic results from Eqs.~(13) and (14), and the solid lines are estimates with $\alpha_\text{eff}$, including the kinetic energy correction in the magnetization at the obstacle's center (Eq.~(18)).}
  \label{fig.3}
\end{figure}

\subsection{Fast Obstacle}

To investigate how the flow patterns evolve with increasing obstacle velocity, we numerically calculated $\{n_i(\bm{r}), S_i(\bm{r})\}$ from Eqs.~(5) and (6) for various $u$. A finite difference method was employed for a $241 \times 241$ grid system, and the obstacle width, $\sigma$, was set to 40 grid spacings. The boundary conditions imposed were $n_i=n_0$ and $\bm{v}_i=-u \hat{\bm{x}}$ on the edge of the grid system. 

Figures~3(a)--(d) show the numerical results for the spin and the mass currents for $u=0.1 c_s$ and $0.9 c_s$ with $V_0=0.1 m c_s^2$. The numerical results for low $u$ were confirmed to be in good quantitative agreement with the analytical predictions based on Eqs.~(13) and (14). We observe that as $u$ increases, the spatial distributions of $\bm{J}$ and $\bm{M}$ stretch along the lateral and the moving directions, respectively, but the flow patterns maintain their characteristic spatial structures~[Figs.~3(c) and (d)]. The peak currents still occur at the center of the moving obstacle. In Figs.~3(e) and (f), we plot $|\bm{J}_0|$ and $|\bm{M}_0|/|\bm{J}_0|$ as functions of $u$, respectively. For low $u$, $|\bm{J}_0|$ increases linearly with $u$, as predicted in Eq.~(13); however, it begins deviating upwardly as $u$ increases over $\approx 0.6 c_s$. The ratio $|\bm{M}_0|/|\bm{J}_0|$ increases quadratically with increasing $u$, departing from the predicted value of $\frac{\ln 2}{2}\alpha$. 
 
The nonlinear $u$-dependence of $\bm{J}_0$ can be attributed to the additional density variations due to the increased flow velocity for high $u$. When the first-order kinetic energy correction term related to $u^2$ in Eq.~(5) is considered, the magnetization can be expressed as 
\begin{equation}
m_z =  \frac{2V}{\Delta g} + \frac{m}{2 \Delta g} \big( 2\bm{u}\cdot(\bm{u}_\uparrow-\bm{u}_\downarrow)\big).
\end{equation}
Because $\bm{u}_{rel}=\bm{u}_\uparrow-\bm{u}_\downarrow \approx \frac{\bm{J}}{n_0}$, the magnetization is enhanced in the center region where the spin current flows in the direction of the obstacle's motion. As the first relation of Eq.~(16) shows, this enhancement in $m_z$ results in an increase in the spin current. This mutual enhancing effect qualitatively explains the observed lateral stretching of the elongated, high-$|\bm{J}|$ region with high $u$.

If the magnetization distribution maintains its Gaussian form for high $u$, i.e., $m_z(r)=m_{z,0} e^{-\frac{2r^2}{\sigma^2}}$, we can infer ${\bm{J}_0}=\frac{m_{z,0}}{2} \bm{u}$ from Eq.~(13) because of the relation $\alpha=\frac{m_{z,0}}{2 n_0}$. Substituting $\bm{u}_{rel}\approx \frac{\bm{J}_0}{n_0}= \frac{m_{z,0}}{2n_0} \bm{u}$ into Eq.~(17), we obtain the magnetization at the obstacle's center as $m_{z,0}=\frac{2 n_0 V_0}{\Delta g n_0-\frac{1}{2}mu^2 }$. This suggests that the high-$u$ effect in the spin and the mass currents might be captured by replacing $\alpha$ in Eqs.~(13) and (14) with its effective value, i.e.,
\begin{equation}
     \alpha_\text{eff}(u)=\frac{m_{z,0}}{2n_0}=\frac{V_0}{\Delta g n_0-\frac{1}{2}mu^2}. 
\end{equation}
In fact, we observe that the numerical results for $|\bm{J}_0|$ and $|\bm{M}_0|/|\bm{J}_0|$ can be explained well quantitatively with $\alpha_\text{eff}$, i.e., $|\bm{J}_0|=\alpha_\text{eff} n_0 u$ and $|\bm{M}_0|/|\bm{J}_0|=\frac{\ln 2}{2} \alpha_\text{eff}$, respectively [Figs.~3(e) and (f)].

\subsection{Critical Velocity}

When the obstacle's velocity increases above a certain critical value, energy dissipation will occur in the binary superfluid system. According to the Landau criterion, the critical velocity is expressed as $u_L=\min [\varepsilon(\bm{p})/(\bm{p}\cdot \hat{\bm{u}})]$~\cite{R1}, where $\varepsilon(\bm{p})$ is the elementary excitation energy of momentum $\bm{p}$, and $\hat{\bm{u}}$ is the unit vector along the direction of the obstacle's motion. In general, in the long wavelength limit, the superfluid system has a linear dispersion of $\varepsilon(p)=c p$ with $c$ being the speed of sound, and the Landau critical velocity is given as $u_L=c$. In this section, we investigate the critical velocity $u_c$ of the magnetic obstacle based on the local Landau criterion, i.e., by comparing the obstacle's velocity to the local speed of sound.

\begin{figure}[t!]
    \includegraphics[width=8.4cm]{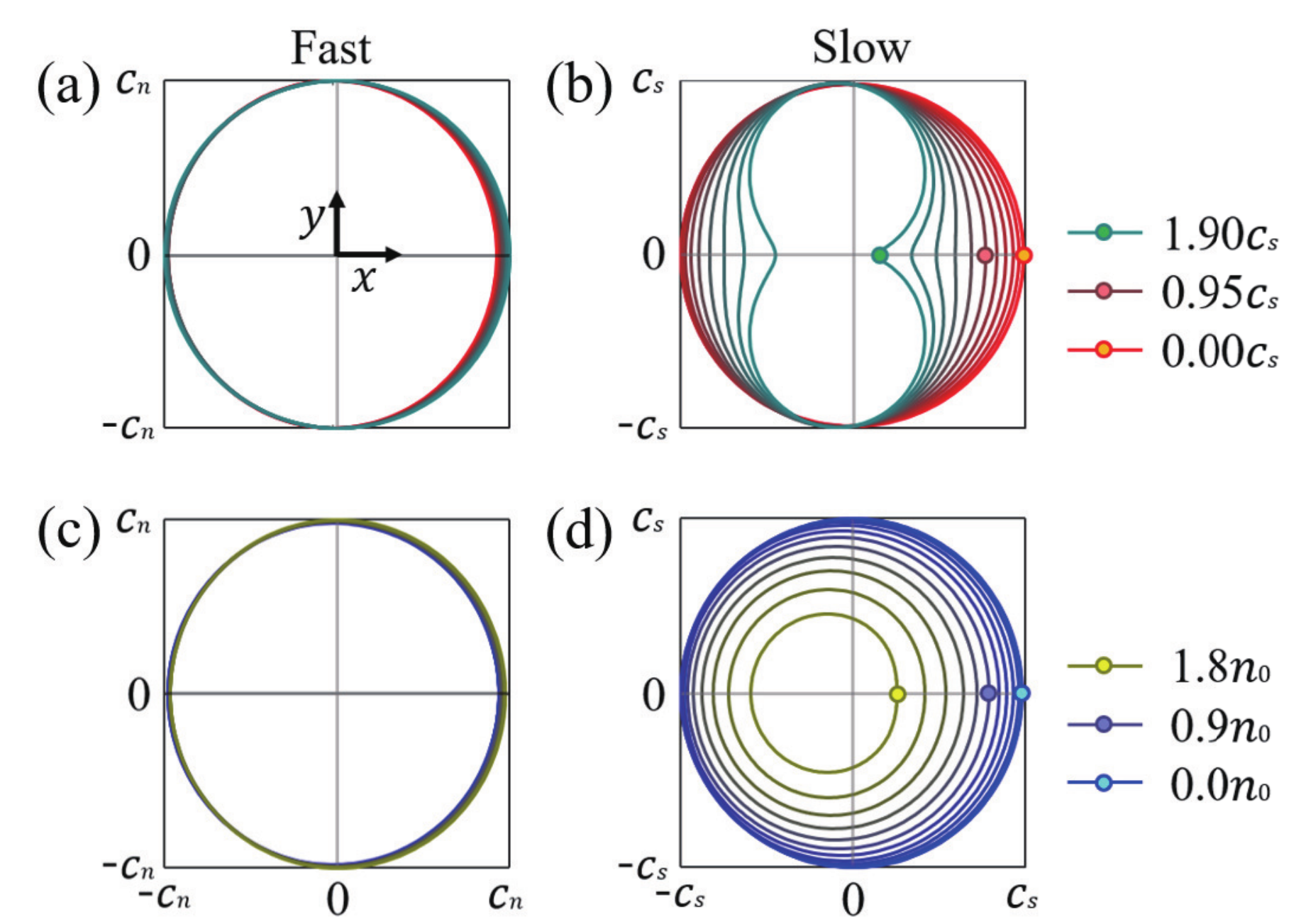}
    \caption{Sound speed in a homogeneous two-component BEC. Radial distributions of the propagation speeds $c^{+}$ and $c^{-}$ of  the fast (a, c) and slow (b, d) sounds, respectively, for various flow conditions of $\bm{u}_{\uparrow(\downarrow)} =\pm \frac{u_{rel}}{2}\hat{\bm{x}}$ and $n_{\uparrow(\downarrow)}=n_0\pm\frac{m_z}{2}$. $\Delta g/g=0.07$ and $c_{n(s)}$ denotes the propagation speed of mass (spin) sound for $u_\text{rel}=0$ and $m_z=0$. In (a) and (b), $m_z=0.36n_0$ and $u_\text{rel}$ changes from 0 to 1.9$c_s$ in intervals of 0.19$c_s$ for the ten lines. In (c) and (d), $u_{\text{rel}}=0.37c_s$ and $m_z$ changes from 0 to 1.8$n_0$ in intervals of 0.18$n_0$ for the ten lines.}
    \label{fig.4}
\end{figure}
 
First, we determine the speed of sound for a stationary state, in which the two spin components flow with uniform velocity $\bm{u}_i$, having unifrom density $n_i$. Linearizing the hydrodynamic equations, Eqs.~(1) and (2), with $V_i=0$~\cite{R28}, we obtain
\begin{eqnarray}
\big (\partial_t +\bm{u}_i \cdot \nabla \big) \delta n_i &+& n_i \nabla \cdot \delta \bm{u}_i =0, \\
\big(\partial_t +\bm{u}_i \cdot \nabla \big) \delta \bm{u}_i &+& \frac{g}{m}\nabla \delta n_i +\frac{g_{\uparrow \downarrow}}{m}\nabla \delta n_j=0.
\end{eqnarray}
Furthermore, the coupled wave equations for $\delta n_\uparrow$ and $\delta n_\downarrow$ can be obtained as follows:
\begin{equation}
\Big( \big(\partial_t +\bm{u}_i \cdot \nabla \big )^2- \frac{g n_i}{m} \nabla^2 \Big ) \delta n_i - \frac{g_{\uparrow\downarrow} n_i}{m} \nabla^2 \delta n_j=0. 
\end{equation}
If a traveling wave solution of $\delta n_i=A_i e^{i (\bm{q}\cdot \bm{r}-\omega t)}$ is to be obtained, the wave velocity $c=\omega/q$ should satisfy 
\begin{equation}
    \frac{A_{\downarrow}}{A_{\uparrow}}=\frac{(c-\bm{u}_\uparrow\cdot \hat{\bm{q}})^2 -\frac{g n_\uparrow}{m}}{\frac{g_{\uparrow\downarrow}n_{\uparrow}}{m}}=\frac{\frac{g_{\uparrow\downarrow}n_{\downarrow}}{m}}{(c-\bm{u}_\downarrow\cdot \hat{\bm{q}})^2 -\frac{g n_\downarrow}{m}},
\end{equation}
with $\hat{\bm{q}}=\frac{\bm{q}}{|\bm{q}|}$. In general, four solutions for $c$ are provided, but because $c(-\hat{\bm{q}})=-c(\hat{\bm{q}})$, we consider only the two positive solutions for the propagation direction of $\hat{\bm{q}}$ and denote them by $c^{+}$ and $c^{-}$ with $c^{+}\geq c^{-}$. For $n_i=n_0$ and $\bm{u}_i=0$, the fast (slow) sound speed is given by $c^{\pm}=c_{n(s)}=\sqrt{\frac{(g \pm g_{\uparrow\downarrow})n_0}{m}}$, and the sound propagates with $A_\uparrow=A_\downarrow$ ($A_\uparrow=-A_\downarrow$), corresponding to phonon (magnon) excitations in a symmetric binary superfluid system.

Figure~4 shows the sound speeds $c^{\pm}(\hat{\bm{q}})$ for various flow conditions of $\bm{u}_\uparrow =\frac{u_{\text{rel}}}{2}\hat{\bm{x}}$, $\bm{u}_\downarrow =-\frac{u_{\text{rel}}}{2}\hat{\bm{x}}$, $n_\uparrow=n_0+\frac{m_z}{2}$, and $n_\downarrow=n_0-\frac{m_z}{2}$. We observe that $c^{+}$ is not significantly affected by changes in $u_{\text{rel}}$ and $m_z$, implying the strong phonon characteristics of the fast sound. Meanwhile, $c^{-}$ is sensitive to them. This decreases with increasing $u_{\text{rel}}$ and $m_z$, and the reduction rate is the fastest along the spin current direction. In our moving-obstacle situation, the relative velocity and the density imbalance between the two spin components are maximum at the obstacle's center. Additionally, the obstacle's direction of motion is the same as the direction of the spin current. Therefore, as the obstacle's velocity increases, the stability of the induced superflow will break first in the region of  the obstacle's center according to the local Landau criterion. 

Figure~5(a) shows the speed of sound, $c^{-}(\hat{\bm{x}})$, at the obstacle's center as a function of $u$ for various $\alpha$ from $0.1$ to $0.9$. The local flow condition of $\{n_i,\bm{u}_i\}$ in the center region was numerically obtained for each set of $\{u, \alpha\}$, and $c^{-}(\hat{\bm{x}})$ was determined from Eq.~(22). As $u$ increases, $c^{-}$ decreases and eventually becomes equivalent to $u$, which defines the obstacle's cirtical velocity, $u_c$. Note that the countersuperflow instability corresponds to an imaginary solution of $c$ in Eq.~(22) and is irrelevant to our current study.

\begin{figure}[t!]
    \includegraphics[width=8.4cm]{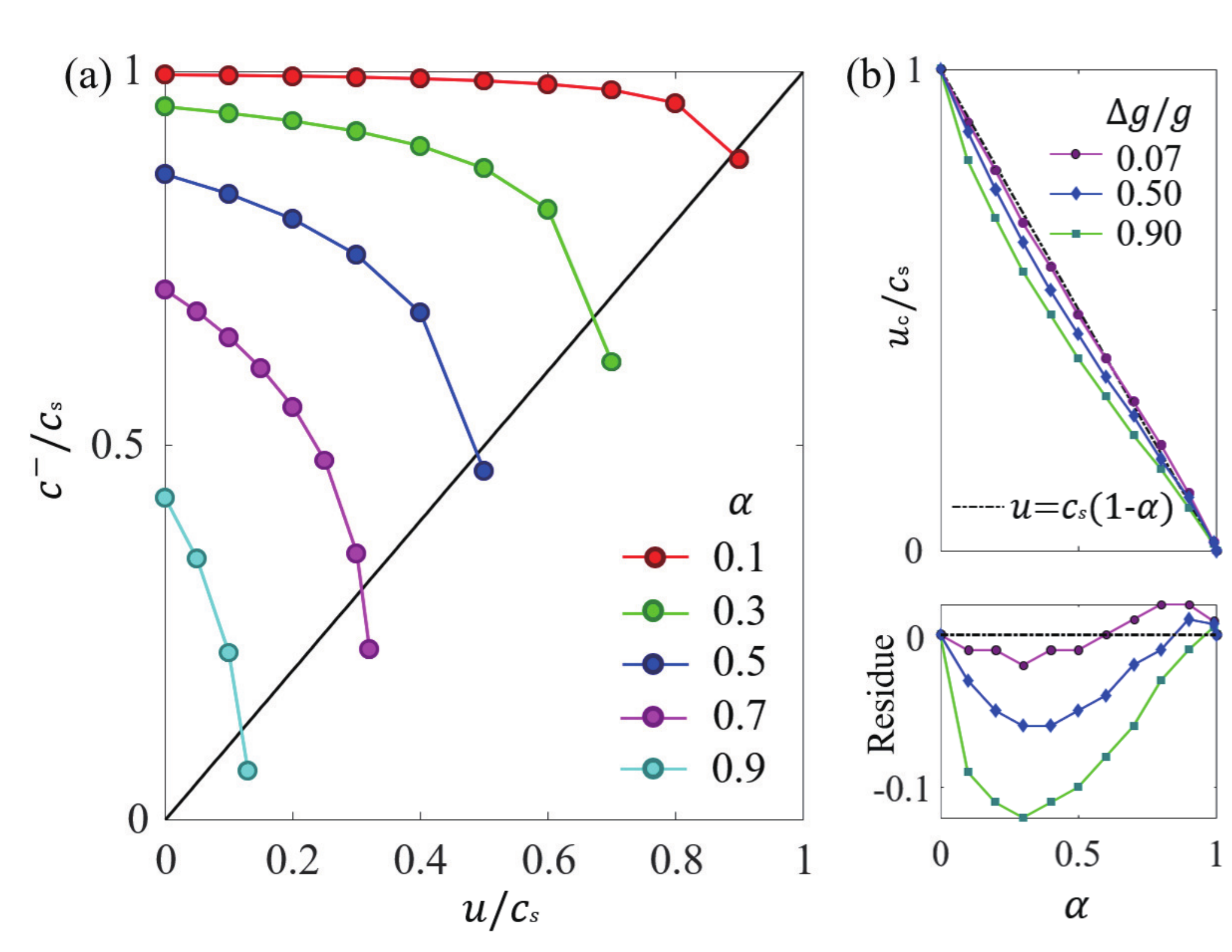}
    \caption{Critical velocity of the magnetic obstacle. (a) Slow sound speed, $c^{-}(\hat{\bm{x}})$, at the obtacle's center as a function of the obstacle velocity $u$ for various potential magnitudes $V_0(=\alpha m c_s^2)$. $\Delta g/g=0.07$. The flow condition of $\{n_{\uparrow(\downarrow)}, \bm{u}_{\uparrow(\downarrow)}\}$ at the obstacle center was numerically determined for given $u$ and $V_0$, and $c^{-}(\hat{\bm{x}})$ was calculated from Eq.~(22). The critical velocity $u_c$ is determined at the onset of Landau instability, where the obstacle's velocity (dashed line) exceeds the sound speed. (b) $u_c$ as a function of $\alpha$ for various $\Delta{g}/g$. The bottom graph shows the residue of $u_c$ with respect to a model critical line of $u_c={c_s}(1-\alpha)$.}
    \label{fig.5}
\end{figure}

The critical velocity $u_c$ decreases with increasing potential magnitude, $V_0$ [Fig.~5(b)], which is attributable to the reduction in $c^-$ due to the increased $m_{z,0}$. In the limit of $V_0\rightarrow 0$, $u_c$ approaches $c_s$ as $m_{z,0}\rightarrow 0$ and $\bm{u}_i\rightarrow 0$. When $V_0$ reaches $m c_s^2$, i.e., $\alpha=1$, $u_c$ vanishes because the system is fully polarized at the obstacle's center and $c^{-}=0$. Interestingly, our numerical results indicate that $u_c$ decreases almost linearly with increasing $V_0$, suggesting an empirical critical line of $u_c(V_0)=c_s (1- \frac{V_0}{m c_s^2})$. We also scrutinized how $u_c$ was affected by the intercomponent interaction strength and observed that when $\Delta{g}/g$ increased from the $^{23}$Na value of 0.07, $u_c$ decreased for the same obstacle condition [Fig.~5(c)]. Because $\Delta{g}/g=1$ corresponds to a non-interacting two-component case, we may conclude that the observed linear dependence of $u_c$ on $V_0$ is driven by the interactions between the two spin components.

\section{Summary and Outlook}
We investigated the spin and the mass flow distributions generated by a moving, penetrable magnetic obstacle in a symmetric binary BEC. We presented an analytical description of the flow patterns in the perturbative regime for a slow obstacle and demonstrated that the induced spin and mass currents exhibit peculiar spatial distributions resembling those of the electric field from a charge dipole and the magnetic field around a current loop, respectively. When the obstacle's velocity was increased, we numerically observed that the spin and the mass flow patterns maintained their overall structures and that the peak current magnitudes were well accounted for by the enhanced spin polarization at the obstacle's center. Finally, we investigated the critical velocity $u_c$ of the magnetic obstacle based on the local Landau instability of the induced superflows and found that $u_c$ almost decreased linearly from the speed of spin sound with the increasing magnitude $V_0$ of the obstacle's potential.

The predicted $u_c(V_0)$ can be immediately tested in current experiments by measuring the rate of temperature increase of a stirred sample as a function of the obstacle's velocity. In previous experiments, the spin temperature of the two-component $^{23}$Na BEC was indirectly probed via the magnitude of spin fluctuations in the sample. When the obstacle's velocity exceeds a critical velocity, the magnetic obstacle will emit magnons, which can be detected as a sudden enhancement in spin fluctuations in the sample. When $u$ is increased further, another critical phenomonon involving the generation of topological objects, such as half-quantum vortices and magnetic solitons, may occur~\cite{Kamchatnov13,R36}. We also notice that another velocity point  larger than $u_c$ exists, above which the obtacle center becomes fully polarized. This may facilitate a phase-slip process in the density-depleted spin component, possibly resulting in vortex nucleation. For a single-component BEC, vortex dipoles were experimentally observed to be periodically generated from a moving, penetrable obstacle~\cite{R15} and that a von Kármán vortex street was formed with an impenetrable obstacle~\cite{R11,R10}.

Finally, we point out that the experimental study with the two-component $^{23}$Na BEC can be extended to spin-1 spinor physics by rendering the $m_F=0$ spin state energetically accessible via tuning the quadratic Zeeman energy. In this case, the spin exchange process of $|m_F=1\rangle+|m_F=-1\rangle \rightarrow 2|m_F=0\rangle$ will be allowed for high spin currents~\cite{R16}, and the critical dynamics with the moving magnetic obstacle is expected to be richer for possibly involving different types of topological defects such as skyrmions~\cite{Choi12}.

\begin{acknowledgments}
We thank Joon Hyun Kim for his discussion and critical reading of the manuscript. This study was supported by the National Research Foundation of Korea (NRF-2018R1A2B3003373, NRF-2019M3E4A1080400). 
\end{acknowledgments}

{}
\end{document}